\documentclass[final,leqno,onefignum,onetabnum]{siamltex1213}

\usepackage{hyperref}
\usepackage{url}
\usepackage{array}
\usepackage{amstext}
\usepackage{amssymb}
\usepackage{amsmath}
\usepackage{amsfonts}
\usepackage{breqn}
\usepackage{graphicx}
\usepackage{subfigure}
\usepackage{wrapfig}
\usepackage{tikz}
\usepackage[plain]{algorithm}
\usepackage{algorithmic}
\usepackage{arydshln}
\usepackage{longtable}
\usepackage{lscape}
\usepackage{multirow}
\usepackage{float}
\usepackage{listings,url,verbatim}
\usepackage{caption}
\usepackage{booktabs}
\usepackage{afterpage}
\newtheorem{prop}[theorem]{Proposition}

\title{An Efficient Algorithm for Unweighted Spectral Graph Sparsification\thanks{This work was partially supported by NSF Award CCF-131912}}

\author{David G. Anderson\thanks{Department of Mathematics, University of California, Berkeley, California 94720 (\email{anderson@math.berkeley.edu}, \email{mgu@math.berkeley.edu}, \email{melgaard@math.berkeley.edu}).} \and Ming Gu\footnotemark[2] \and Christopher Melgaard\footnotemark[2]}

\begin{document}
\maketitle
\slugger{sicomp}{xxxx}{xx}{x}{x--x}

\begin{abstract}
Spectral graph sparsification has emerged as a powerful
tool in the analysis of large-scale networks by reducing the overall
number of edges, while maintaining a comparable graph Laplacian
matrix. In this paper, we present an efficient algorithm for the
construction of a new type of spectral sparsifier, the
\emph{unweighted} spectral sparsifier. Given a general undirected and unweighted
graph $G = \left( V, E \right)$, and an integer $\ell < |E|$ (the
number of edges in $E$), we compute an unweighted graph $H = \left( V,
F \right)$ with $F \subset E$ and $|F| = \ell$ such that for every $x \in \mathbb{R}^{V}$ 
\[
{\displaystyle \frac{x^T L_G x}{\kappa} \leq x^T L_H x \leq x^T L_G
x,} \] 
where $L_G$ and $L_H$ are the Laplacian matrices for $G$ and $H$,
respectively, and $\kappa \geq 1$ is a slowly-varying
function of $|V|, |E|$ and $\ell$. This work addresses the
open question of the existence of \emph{unweighted} graph
sparsifiers for unweighted
graphs~\cite{ramanujansparse}.  Additionally, our algorithm efficiently computes
unweighted graph sparsifiers for weighted graphs, leading to
sparsified graphs that retain the weights of the original graphs.
\end{abstract}



\pagestyle{myheadings}
\thispagestyle{plain}
\markboth{AN ALGORITHM FOR UNWEIGHTED GRAPH SPARSIFICATION}{D. G. ANDERSON, M. GU, AND C. MELGAARD}

\section{Introduction}

Graph sparsification seeks to approximate a graph $G$ with a graph $H$ on the same vertices, but with fewer edges.  Called a sparsifier, $H$ requires less storage than $G$ and serves as a proxy for $G$ in computations where $G$ is too large, evoking the effectiveness of sparsifiers in wide-ranging applications of graphs, including social networks, conductance, electrical networks, and similarity \cite{conf/soda/ChierichettiLP10,DBLP:conf/stoc/ChristianoKMST11,conf/kdd/MathioudakisBCGU11,st1}.  In some applications, graph sparsification improves the quality of the graph, such as in the design of information networks and the hardwiring of processors and memory in parallel computers \cite{journals/cacm/BatsonSST13,4495}.  Sparsifiers have also been utilized to find approximate solutions of symmetric, diagonally-dominant linear systems in nearly-linear time \cite{journals/cacm/BatsonSST13,journals/corr/cs-DS-0310036,st1,journals/corr/abs-cs-0607105}.

Recent work on graph sparsification includes \cite{agm,conf/innovations/KapralovP12,ss1,st1,spectsim}.  Batson, Spielman, and Srivastava \cite{ramanujansparse} prove that for every graph there exists a spectral sparsifier where the number of edges is linear in the number of vertices.  They further provide a polynomial-time, deterministic algorithm for the sparsification of weighted graphs, which could produce weights that differ greatly from the weights of the original graph.  The work of Avron and Boutsidis \cite{fasterSub} explores unweighted sparsification in the context of finding low-stretch spanning trees.  They provide a greedy edge removal algorithm and a volume sampling algorithm with theoretical guarantees.  In comparison, our novel greedy edge selection algorithm has tighter theoretical bounds for both spanning trees and in the more general context of unweighted graph sparsification.

Our work introduces a deterministic, greedy edge selection algorithm to calculate sparsifiers for weighted and unweighted graphs.  Our algorithm selects a subset of edges for the sparse approximation $H$, without assigning or altering weights.  While the Dual Set algorithms of \cite{fasterSub, ramanujansparse, nearoptcol} reweight all selected edges for computing \emph{weighted sparsifiers}, our algorithm produces \emph{unweighted sparsifiers} for an unweighted input graph, and can create a weighted sparsifier for a weighted input graph by assigning the original edge weights to the sparsifier.  Hence our concept of unweighted sparsification applies to both unweighted and weighted graphs.  To formalize:

\begin{definition} 
Let $G = \left(V,E, w \right)$ be a given graph\footnote{Note that any unweighted graph $G =
\left(V,E\right)$ induces a weighted graph $G = \left( V, E,
w \right)$ where $w_{e} = 1$ if $e=(u,v)\in E$ and
$w_{e} = 0$ otherwise.
}. We define an
{\bf unweighted sparsification} of $G$ to be any graph of the form $H
= \left(V, F,w \odot \mathbb{I}_F \right)$, where 
\[\mathbb{I}_F(e) = \left\{ \begin{array} {ll} 1, & \text{ if } e \in F \\ 0, &
\text{ otherwise} \end{array} \right. \]
is the indicator function and
$\odot$ is the Hadamard product, i.e.
\[ \left(w \odot \mathbb{I}_F \right) (e) = \left\{
\begin{array}{ll} w_{e}, & \text{ if } e = (u,v) \in F \\ 0, &
\text{ otherwise} \end{array}. \right.\]
\end{definition}

Several definitions have been proposed for the notion in which a sparsifier approximates a dense graph.  Benzc\'{u}r and Karger \cite{conf/stoc/BenczurK96} introduced cut sparsification, where the sum of the weights of the edges of a cut dividing the set of vertices is approximately the same for the dense graph and the sparsifier.  Spielman and Teng \cite{spectsim} proposed spectral sparsification, a generalization of cut sparsification, which seeks sparsifiers with a Laplacian matrix close to that of the input graph.  We follow the work of \cite{ramanujansparse,spectsim} and base our work on spectral sparsification, for which we now present a rigorous definition.

Given an undirected graph $G= \left( V, E, w \right)$, define the signed edge-vertex incidence matrix $B_G \in \mathbb{R}^{E \times V }$ as
\[ \left( B_G \right)_{ej} = \left\{ \begin{array}{ll} -1, & \text{ if
$e = (u,v) \in E$ and $j=u\in V$} \\ 1, & \text{ if $e=(u,v)\in E$ and
$j=v\in V$,} \\ 0 & \text{ otherwise} \end{array} \right. \]
where all edges are randomly assigned a direction, and $e = \left( u, v \right) \in E$ is an edge from $u$ to $v$.  Define the diagonal weight matrix $W_G \in \mathbb{R}^{E \times E}$
\[ \left(W_G \right)_{ef} = \left\{ \begin{array}{ll} w_e, &
\text{ if $e=f\in E$}, \\ 0 & \text{ otherwise} \end{array}. \right.
\]
The Laplacian of the graph is
\[ L_G = B_G^T W_G B_G .\]
Note that 
\[ x^T L_G x = \sum_{(u,v) \in E} w_{(u,v)} \left( x_u - x_v
\right)^2 \]
for a vector $x \in \mathbb{R}^{ V }$.  To compare Laplacians of graphs $X$ and $Y$ defined
on the same set of nodes we denote
\[L_X \preceq L_Y \hspace{0.6cm} \text{ if and only if } \hspace{0.6cm}  x^T L_X x \leq x^T L_Y x \text{, } \hspace{0.2cm} \text{ for all } x. \]
\begin{definition} 
The graph $H$ is a {\bf $\kappa$-approximation} of $G$ if
\[ \frac{1}{\kappa} L_G \preceq L_H \preceq L_G. \]
\end{definition}
Because our unweighted sparsification algorithm does not change the weights of
the edges kept in $H$, it is immediate that $L_H \preceq L_G$:
\begin{prop} \label{prop:upp}
If $H$ is an unweighted sparsification of $G$, then
\[L_H \preceq L_G. \]
\proof
\begin{eqnarray}
x^T L_H x &=& \sum_{(u,v) \in F} w_{(u,v)} \left( x_u - x_v\right)^2\nonumber\\
&\leq& \sum_{(u,v) \in F} w_{(u,v)} \left( x_u - x_v\right)^2 + \sum_{(u,v) \in E \setminus F} w_{(u,v)} \left( x_u -
x_v \right)^2\nonumber\\
&=&x^T L_G x\nonumber
\end{eqnarray}
for all $x \in \mathbb{R}^{V}$.
\hspace{2 em}\endproof
\end{prop}

Our algorithm does not operate directly on the Laplacian matrix.  Rather, we consider the SVD of $W_G^{1/2} B_G$.

\begin{equation}\label{PolarMat}
W_G^{1/2} B_G = U_G^T \Sigma_G V_G, 
\end{equation}

\noindent where $\Sigma_G$ is a diagonal matrix containing all non-zero singular values of $W_G^{1/2} B_G$; and where $U_G \in \mathbb{R}^{n\times m}$ is a row orthonormal matrix, with $n = |V|-r$, and $r$ being the number of connected components in $G$. For the unweighted graph, $W_G$ is simply the
identity matrix. $U_G$ plays a similar role to that of the matrix
$V_{n\times m}$ in~\cite{ramanujansparse} and the matrix $Y$ in \cite{fasterSub}.  Our algorithm utilizes the column-orthogonality of $U_G^T$, highlighting the reason for not working directly with the Laplacian matrix.  We note, nevertheless, that this algorithm can be adapted to any orthogonal decomposition of $W_G^{1/2} B_G$.

We are now in a position to present our main results.  In section 2 we present the unweighted column selection algorithm, as well as the spectral bounds for the sparsifiers it calculates.  Section 3 provides supporting theory.  Comparisons to other modern algorithms are made in section 4.  In section 5 we demonstrate one application of graph sparsification, graph visualization, by applying our algorithm to real autonomous systems data.  Some observations and concluding remarks are offered in sections 6 and 7.

\section{The Unweighted Column Selection (UCS) Algorithm}
Our algorithm selects edges for a sparsifier based on the columns ${u}_i$ of $U_G$,
\begin{eqnarray}U_G=\left(\begin{array}{cccc}u_1 & u_2 & \cdots & u_m\end{array}\right)\in\mathbb{R}^{n\times m},\nonumber\end{eqnarray}
where $m = \left| E \right|$ is the number of edges, and $n = \left| V
\right|-r$, as above. Therefore, the edges of $G$ that
are included in the sparsifier are exactly the columns of $U_G$ that our algorithm selects. Denote
the number of edges kept as $\ell \stackrel{def}= \left| F \right|$.  Let $\Pi_t$ denote the set of selected edges after $t$ iterations.

We propose the following greedy algorithm for column selection on $U_G$.  Initially set $A_0=0_{n \times n}$ and $\Pi_0=\varnothing$, and
choose a constant $T>0$. At step $t \geq 0$:
\begin{itemize}
\item Solve for the unique $\lambda < \lambda_{\min}\left(A_t\right)$ such that
\begin{eqnarray}
\text{\bf tr}\left(A_t-\lambda I\right)^{-1}=T . \label{step1}
\end{eqnarray}
\item Solve for the unique $\widehat{\lambda} \in \left(\lambda,
\lambda_{\min}\left(A\right)\right)$ such that
\begin{eqnarray}
{\displaystyle \left(\widehat{\lambda}-\lambda\right)\left(m-t+\sum_{j=1}^n\frac{1-\lambda_j}{\lambda_j-\lambda}\right)} ={\displaystyle \frac{\sum_{j=1}^n\frac{1-\lambda_j}{\left(\lambda_j-\lambda\right)\left(\lambda_j-\widehat{\lambda}\right)}}{\sum_{j=1}^n\frac{1}{\left(\lambda_j-\lambda\right)\left(\lambda_j-\widehat{\lambda}\right)}} ,}\label{step2}
\end{eqnarray}
where $\lambda_j$ is the $j^{th}$ largest eigenvalue of the symmetric matrix $A_t$.
\item Find an index $i\notin\Pi_t$ such that 
\begin{eqnarray}
{\displaystyle \text{\bf tr}\left(A_t-\widehat{\lambda}I+{u}_i{u}_i^T\right)^{-1}} \leq {\displaystyle \text{\bf tr}\left(A_t-\lambda I\right)^{-1}.} \label{step3}
\end{eqnarray}
\item Update $A_t$ and $\Pi_t$. 
\end{itemize}

While equations (\ref{step1}) and (\ref{step2}) are relatively straightforward to justify and solve, equation (\ref{step3}) requires careful consideration, and is the focus of much of section 3.  Note that equation ($\ref{step1}$) can be solved in $O\left(n^3\right)$ operations, equation ($\ref{step2}$) in $O\left(n\right)$ operations, and equation ($\ref{step3}$) in $O\left(n^2m\right)$ operations.  This last complexity count follows because testing the inequality scales with $O\left(n^2\right)$, and potentially all remaining indices must be tested.  Thus the total complexity of selecting $\ell$ columns is $O\left(\ell n^2m\right)$.

While this procedure will work for any $T > 0$, we will show that an effective choice is
\begin{equation}
T=\widehat{T}^*\left(1+F\left(\widehat{T}^*\right)\right),\nonumber
\end{equation}
where 
$$F\left(\widehat{T}\right)=\left[\left(1-\frac{n}{\widehat{T}}\right)\frac{\ell}{m-\frac{\ell-1}{2}+\widehat{T}-n}-\frac{n}{\widehat{T}}\right],$$
and where $\widehat{T}^*$ is the minimizer of $F\left(\widehat{T}\right)$, given as
$$\widehat{T}^*=\frac{n\left(m+\frac{\ell+1}{2}-n\right)+\sqrt{n\ell\left(m-\frac{\ell-1}{2}\right)\left(m+\frac{\ell+1}{2}-n\right)}}{\ell-n}.$$
Our spectral bounds are derived using this choice of $T$.  We summarize this procedure in the Unweighted Column Selection algorithm.

\begin{algorithm}
{\sc Algorithm: Unweighted Column Selection (UCS)}

\renewcommand{\algorithmicrequire}{\textbf{Inputs:}}
\renewcommand{\algorithmicensure}{\textbf{Outputs:}}
\renewcommand{\algorithmicprint}{\textbf{break}}
    \refstepcounter{algorithm}
    \label{algo:unweighted}
    \begin{algorithmic}[1]
    \REQUIRE $G=\left( V, E, w \right)$, $T > 0$, $\ell$.
    \ENSURE $H_{uw} = \left( V, F, w \odot \mathbb{I}_F \right)$
    \STATE Calculate the column-orthogonal matrix $U_G^T$
    \STATE Set $A_0=0_{n \times n}$, $\Pi_0=\varnothing$
    \FOR{$t=0,\cdots,\ell-1$}
        \STATE Solve for $\lambda$ using equation ($\ref{step1}$)
        \STATE Calculate $\widehat{\lambda}$ using equation ($\ref{step2}$)
        \STATE Find $i\not\in \Pi_t$ such that inequality ($\ref{step3}$) is satisfied \label{find_i}
        \STATE Update $A_{t+1}=A_t+{u}_i{u}_i^T$
        \STATE Update $\Pi_{t+1}=\Pi_t\cup\{i\}$
    \ENDFOR
		\STATE Let $F=\Pi_{\ell}$ be the selected edges
    \end{algorithmic}
\end{algorithm}

Theorem~\ref{thm:unweighted} below confirms the correctness of the Unweighted Column Selection Algorithm. This theorem, along with other properties of 
the UCS algorithm, will be discussed and proved in Section~\ref{Sec:Properties}.
\begin{theorem} \label{thm:unweighted}
Let $G=\left(V,E,w\right)$ and let $n < \ell < m$. Then the sparsified graph $H$ produced by 
the UCS algorithm satisfies
\begin{equation}
\frac{1}{\kappa} L_G \preceq L_H \preceq L_G ,\nonumber
\end{equation}
where 
\begin{equation}\label{Eqn:alpha}
\frac{1}{\kappa} = \frac{\left(\ell-n\right)^2}{{\left(\sqrt{n\left(m+\frac{\ell+1}{2}-n\right)}+\sqrt{\ell\left(m-\frac{\ell+1}{2}\right)}\right)^2+(\ell-n)^2}}.
\end{equation}
\end{theorem}

\section{Correctness and Performance of the UCS Algorithm}\label{Sec:Properties}
The goal of this section is to prove Theorem~\ref{thm:unweighted}.  Section \ref{sec:31} establishes that the UCS algorithm is well-defined.  Section \ref{sec:32} proves a lower bound for the minimum singular value of the submatrix selected by the UCS algorithm, and provides a good choice for the input parameter $T$.  In section \ref{sec:33}, the UCS algorithm is shown to be a graph sparsification algorithm.

\subsection{The Existence of a Solution to Equation~(\ref{step3})}\label{sec:31}
The next two lemmas show that equation~(\ref{step3}) always has a solution.
\begin{lemma}\label{lemma:1}
At a given iteration $t$ in the UCS algorithm, at step $\ref{find_i}$ define
\begin{eqnarray}
f(x)&\stackrel{def}=&{\displaystyle \left(x-\lambda\right)\left[m-t+\sum_{j=1}^n\frac{1-\lambda_j}{\lambda_j-\lambda}\right]-\frac{\sum_{j=1}^n\frac{1-\lambda_j}{\left(\lambda_j-\lambda\right)\left(\lambda_j-x\right)}}{\sum_{j=1}^n\frac{1}{\left(\lambda_j-\lambda\right)\left(\lambda_j-x\right)}}.}\nonumber
\end{eqnarray}
Then there exists $\widehat{\lambda}$, with $\lambda<\widehat{\lambda}<\lambda_n$, such that $f\left(\widehat{\lambda}\right)=0$.  Furthermore,
\small
\begin{eqnarray}
0&<&{\displaystyle \left(\widehat{\lambda}-\lambda\right)\left[\frac{\sum_{j=1}^n\frac{1-\lambda_j}{\left(\lambda_j-\lambda\right)\left(\lambda_j-\widehat{\lambda}\right)^2}}{\sum_{j=1}^n\frac{1}{\left(\lambda_j-\lambda\right)\left(\lambda_j-\widehat{\lambda}\right)}}-\left(\widehat{\lambda}-\lambda\right)\sum_{j=1}^n\frac{1-\lambda_j}{\left(\lambda_j-\lambda\right)\left(\lambda_j-\widehat{\lambda}\right)}\right].}\label{eqn:gz}
\end{eqnarray}
\end{lemma}
\normalsize
\proof
Clearly $f\left(\lambda\right)<0$.  Although $f$ is undefined at
$\lambda_n$, let $\lambda_n^\epsilon:=\lambda_n-\epsilon$, where
$\epsilon > 0$.  Note that
\begin{eqnarray*}
{\displaystyle \lim_{\epsilon\rightarrow 0+}\left(\sum_{j=1}^n\frac{1-\lambda_j}{\left(\lambda_j-\lambda\right)\left(\lambda_j-\lambda_n^\epsilon\right)}\right) \left/ \left(\sum_{j=1}^n\frac{1}{\left(\lambda_j-\lambda\right)\left(\lambda_j-\lambda_n^\epsilon\right)}\right)\right.}  ={\displaystyle 1-\lambda_n}
\end{eqnarray*}
because the last term in each sum will dominate the rest of the sum.  Furthermore,
\begin{eqnarray}
{\displaystyle \lim_{\epsilon\rightarrow 0^+}\left(\lambda_n^\epsilon-\lambda\right)\left[m-t+\sum_{j=1}^n\frac{1-\lambda_j}{\lambda_j-\lambda}\right]}&=&{\displaystyle 1-\lambda_n+\lim_{\epsilon\rightarrow 0^+}\left(\lambda_n^\epsilon-\lambda\right)\left[m-t+\sum_{j=1}^{n-1}\frac{1-\lambda_j}{\lambda_j-\lambda}\right]}\nonumber\\
&>&{\displaystyle 1-\lambda_n }.\nonumber
\end{eqnarray}
Hence for small $\epsilon>0$, we have $f\left(\lambda_n^\epsilon\right)>0$, and, therefore, $\widehat{\lambda}$ exists, with $\lambda<\widehat{\lambda}<\lambda_n$, and $f\left(\widehat{\lambda}\right)=0$ via the Intermediate Value Theorem.  Note that if there exists $0<\gamma<n$ such that $\lambda_{\gamma}=\lambda_{\gamma+1}=\cdots=\lambda_n$, then we repeat the same argument replacing the expression $1-\lambda_n$ with $\sum_{j=\gamma}^n1-\lambda_j=(n-\gamma+1)\left(1-\lambda_n\right)$.\\

Now we prove inequality (\ref{eqn:gz}).  We use the following version of the Cauchy-Schwartz formula: for $a_j,b_j\ge0$ then $\left(\sum a_jb_j\right)^2\le\left(\sum a_j^2 b_j\right)\left(\sum b_j\right)$. Consequently
\begin{eqnarray*}
&&{\displaystyle \left(\sum_{j=1}^n\frac{1-\lambda_j}{\left(\lambda_j-\lambda\right)\left(\lambda_j-\widehat{\lambda}\right)}\right)^2 } \\
&&{\displaystyle  \hspace{1cm}\le\left( \sum_{j=1}^n\frac{1-\lambda_j}{\left(\lambda_j-\lambda\right)\left(\lambda_j-\widehat{\lambda}\right)^2}\right)\left( 0 + \sum_{j=1}^n\frac{1-\lambda_j}{\lambda_j-\lambda}\right)} \\
&&{\displaystyle\hspace{1cm} <\left(\sum_{j=1}^n\frac{1-\lambda_j}{\left(\lambda_j-\lambda\right)\left(\lambda_j-\widehat{\lambda}\right)^2}\right)\left(\overbrace{m-t}^{>0}+\sum_{j=1}^n\frac{1-\lambda_j}{\lambda_j-\lambda}\right)} \\
&&{\displaystyle \hspace{1cm}=\frac{1}{\left(\widehat{\lambda}-\lambda\right)} \left( \sum_{j=1}^n\frac{1-\lambda_j}{\left(\lambda_j-\lambda\right)\left(\lambda_j-\widehat{\lambda}\right)^2}\right)\left(\frac{ \sum_{j=1}^n\frac{1-\lambda_j}{\left(\lambda_j-\lambda\right)\left(\lambda_j-\widehat{\lambda}\right)}}{\sum_{j=1}^n\frac{1}{\left(\lambda_j-\lambda\right)\left(\lambda_j-\widehat{\lambda}\right)}}\right),}
\end{eqnarray*}
where the last step comes from 
$f\left(\widehat{\lambda}\right)=0$. The strict inequality above holds
because $m-t \geq m-\ell+1 \geq 1.$ After some simple algebra,
\small
\[{\displaystyle
\left(\widehat{\lambda}-\lambda\right)\sum_{j=1}^n\frac{1-\lambda_j}{\left(\lambda_j-\lambda\right)\left(\lambda_j-\widehat{\lambda}\right)}< \left(\sum_{j=1}^n\frac{1-\lambda_j}{\left(\lambda_j-\lambda\right)\left(\lambda_j-\widehat{\lambda}\right)^2}\right) \left/ \left(\sum_{j=1}^n\frac{1}{\left(\lambda_j-\lambda\right)\left(\lambda_j-\widehat{\lambda}\right)}\right)\right.,}\]
\normalsize
which implies our desired inequality because $0 < \widehat{\lambda}-\lambda$.
\hspace{2 em}\endproof

Next, we show that our algorithm is well defined in the sense we can always find a new index $i\notin \Pi_t$ for each iteration that
satisfies $(\ref{find_i})$.
\begin{lemma}\label{lemma:2}
An index $i\notin \Pi_t$ can always be found to satisfy line
$(\ref{find_i})$ of the UCS algorithm for $0\leq t
<\ell$.
\end{lemma}
\proof Note the two following partial fraction results
\begin{eqnarray}
{\displaystyle \frac{\widehat{\lambda}-\lambda}{\left(\lambda_j-\widehat{\lambda}\right)\left(\lambda_j-\lambda\right)}}&=&{\displaystyle \frac{1}{\lambda_j-\widehat{\lambda}}-\frac{1}{\lambda_j-\lambda} }\label{eqn1} \\
{\displaystyle \frac{\widehat{\lambda}-\lambda}{\left(\lambda_j-\widehat{\lambda}\right)\left(\lambda_j-\lambda\right)^2}+\frac{1}{\left(\lambda_j-\widehat{\lambda}\right)\left(\lambda_j-\lambda\right)}}&=&{\displaystyle \frac{1}{\left(\lambda_j-\widehat{\lambda}\right)^2}.} \label{eqn2}
\end{eqnarray}

Using the fact that $f(\widehat{\lambda})=0$, followed by the inequality of Lemma $\ref{lemma:1}$, we have
\footnotesize
\begin{eqnarray*}
&&{\displaystyle \left(\widehat{\lambda}-\lambda\right)\left[m-t+\sum_{j=1}^n\frac{1-\lambda_j}{\lambda_j-\lambda}\right] }\\
&&\hspace{1cm}= \left(\sum_{j=1}^n\frac{1-\lambda_j}{\left(\lambda_j-\widehat{\lambda}\right)\left(\lambda_j-\lambda\right)}\right) \left/
\left(\sum_{j=1}^n\frac{1}{\left(\lambda_j-\widehat{\lambda}\right)\left(\lambda_j-\lambda\right)}
\right)\right. + 0 \\
&&{\displaystyle  \hspace{1cm} < \left(\sum_{j=1}^n\frac{1-\lambda_j}{\left(\lambda_j-\widehat{\lambda}\right)\left(\lambda_j-\lambda\right)}\right) \left/ \left(\sum_{j=1}^n\frac{1}{\left(\lambda_j-\widehat{\lambda}\right)\left(\lambda_j-\lambda\right)}\right) \right. }\\
&&{\displaystyle  \hspace{1.5cm} + \left(\widehat{\lambda}-\lambda\right)\left[\frac{\sum_{j=1}^n\frac{1-\lambda_j}{\left(\lambda_j-\widehat{\lambda}\right)\left(\lambda_j-\lambda\right)^2}}{\sum_{j=1}^n\frac{1}{\left(\lambda_j-\widehat{\lambda}\right)\left(\lambda_j-\lambda\right)}}-\left(\widehat{\lambda}-\lambda\right)\sum_{j=1}^n\frac{1-\lambda_j}{\left(\lambda_j-\widehat{\lambda}\right)\left(\lambda_j-\lambda\right)}\right]} \\
&&{\displaystyle  \hspace{1cm} = \left(\left(\widehat{\lambda}-\lambda\right) \sum_{j=1}^n\frac{1-\lambda_j}{\left(\lambda_j-\widehat{\lambda}\right)\left(\lambda_j-\lambda\right)^2} + \sum_{j=1}^n\frac{1-\lambda_j}{\left(\lambda_j-\widehat{\lambda}\right)\left(\lambda_j-\lambda\right)}\right)  }\\
&&{\displaystyle \hspace{1.5cm} \left/
\left(\sum_{j=1}^n\frac{1}{\left(\lambda_j-\widehat{\lambda}\right)\left(\lambda_j-\lambda\right)}\right)\right. - \left( \widehat{\lambda}-\lambda \right)^2 \sum_{j=1}^n\frac{1-\lambda_j}{\left(\lambda_j-\widehat{\lambda}\right)\left(\lambda_j-\lambda\right)}} \\
&&{\displaystyle  \hspace{1cm} = \frac{\sum_{j=1}^n\frac{1-\lambda_j}{\left(\lambda_j-\widehat{\lambda}\right)^2}} 
{\sum_{j=1}^n\frac{1}{\left(\lambda_j-\widehat{\lambda}\right)\left(\lambda_j-\lambda\right)}} - \left( \widehat{\lambda}-\lambda \right) \left(\sum_{j=1}^n\frac{1-\lambda_j}{\lambda_j-\widehat{\lambda}}-\sum_{j=1}^n\frac{1-\lambda_j}{\lambda_j-\lambda}\right),}
\end{eqnarray*} 	
\normalsize
where the last line follows from equations (\ref{eqn1}) and (\ref{eqn2}).  After some rearranging:
\begin{eqnarray}
{\displaystyle  \left(m-t+\sum_{j=1}^n\frac{1-\lambda_j}{\lambda_j-\widehat{\lambda}}\right)\left(\sum_{j=1}^n\frac{\widehat{\lambda}-\lambda}{\left(\lambda_j-\widehat{\lambda}\right)\left(\lambda_j-\lambda\right)}\right) }<{\displaystyle   \sum_{j=1}^n\frac{1-\lambda_j}{\left(\lambda_j-\widehat{\lambda}\right)^2}.}\nonumber
\end{eqnarray}

This inequality can be rewritten using the trace property $\displaystyle \text{\bf tr}\left(xy^T\right)=y^Tx$ and the identity $\displaystyle \sum_{i\notin \Pi_t} u_iu_i^T=\sum_{i=1}^{m} u_iu_i^T - \sum_{i\in \Pi_t} u_iu_i^T = I_n-A_t$:
\footnotesize
\begin{eqnarray*}
&& {\displaystyle \left(\sum_{i\not\in \Pi_t} 1+u_i^T\left(A_t-\widehat{\lambda}I\right)^{-1}u_i\right)\left(\text{\bf tr}\left(A_t-\widehat{\lambda}I\right)^{-1}-\text{\bf tr}\left(A_t-\lambda I\right)^{-1}\right)} \\
&&{\displaystyle   \hspace{1cm} = \left(m-t+\sum_{i\not\in \Pi_t}\text{\bf tr}\left[\left(A_t-\widehat{\lambda}I\right)^{-1}u_iu_i^T\right]\right)\left(\sum_{j=1}^n\frac{1}{\lambda_j-\widehat{\lambda}}-\sum_{j=1}^n\frac{1}{\lambda_j-\lambda}\right) }\\
&&{\displaystyle   \hspace{1cm} = \left(m-t+\text{\bf tr}\left[\left(A_t-\widehat{\lambda}I\right)^{-1}\left(I-A_t\right)\right]\right)\sum_{j=1}^n\frac{\widehat{\lambda}-\lambda}{\left(\lambda_j-\widehat{\lambda}\right)\left(\lambda_j-\lambda\right)} }\\
&&{\displaystyle   \hspace{1cm} = \left(m-t+\sum_{j=1}^n\frac{1-\lambda_j}{\lambda_j-\widehat{\lambda}}\right)\left(\sum_{j=1}^n\frac{\widehat{\lambda}-\lambda}{\left(\lambda_j-\widehat{\lambda}\right)\left(\lambda_j-\lambda\right)}\right)} \\
&& {\displaystyle  \hspace{1cm} < \sum_{j=1}^n\frac{1-\lambda_j}{\left(\lambda_j-\widehat{\lambda}\right)^2}}\\
&& {\displaystyle  \hspace{1cm} = \text{\bf tr}\left(\left(A_t-\widehat{\lambda}I\right)^{-2}\left(I-A_t\right)\right)}\\
&& {\displaystyle  \hspace{1cm} =  \sum_{i\not\in \Pi_t}u_i^T\left(A_t-\widehat{\lambda}I\right)^{-2}u_i.}
\end{eqnarray*}
\normalsize
Moving terms to the right and dividing by ${\displaystyle  \left(\text{\bf tr}\left(A_t-\widehat{\lambda}I\right)^{-1}-\text{\bf tr}\left(A_t-\lambda I\right)^{-1}\right) > 0}$ (because $\widehat{\lambda}>\lambda$)  gives
\[{\displaystyle  \sum_{i\not\in \Pi_t}\left(\frac{u_i^T\left(A_t-\widehat{\lambda}I\right)^{-2}u_i}{\text{\bf tr}\left(A_t-\widehat{\lambda}I\right)^{-1}-\text{\bf tr}\left(A_t-\lambda I\right)^{-1}}-\left(1+u_i^T\left(A_t-\widehat{\lambda}I\right)^{-1}u_i\right)\right) > 0.} \]
For this to be true, there must exist an $i\not\in \Pi_t$ such that
\[{\displaystyle  \left(\frac{u_i^T\left(A_t-\widehat{\lambda}I\right)^{-2}u_i}{\text{\bf tr}\left(A_t-\widehat{\lambda}I\right)^{-1}-\text{\bf tr}\left(A_t-\lambda I\right)^{-1}}-\left(u_i^T\left(A_t-\widehat{\lambda}I\right)^{-1}u_i\right)\right) > 1.} \]
This last relation gives
\begin{eqnarray*}
{\displaystyle  \text{\bf tr}\left(A_t-\lambda I\right)^{-1}} &>&{\displaystyle   \text{\bf tr}\left(A_t-\widehat{\lambda}I\right)^{-1}-\frac{u_i^T\left(A_t-\widehat{\lambda}I\right)^{-2}u_i}{1+u_i^T\left(A_t-\widehat{\lambda}I\right)^{-1}u_i}} \\
&=&{\displaystyle  \text{\bf tr}\left(A_t-\widehat{\lambda}I\right)^{-1}-\text{\bf tr}\left(\frac{\left(A_t-\widehat{\lambda}I\right)^{-1}u_iu_i^T\left(A_t-\widehat{\lambda}I\right)^{-1}}{1+u_i^T\left(A_t-\widehat{\lambda}I\right)^{-1}u_i}\right)} \\
&=&{\displaystyle  \text{\bf tr}\left(A_t-\widehat{\lambda}I+u_iu_i^T\right)^{-1}, }
\end{eqnarray*}
where the last line was accomplished with the trace property previously indicated and the Sherman-Morrison formula.
\hspace{2 em}\endproof

\subsection{Lower Bound on $\lambda_{\min}(A_{\ell})$}\label{sec:32}

Lemma~\ref{lemma:2} ensures that the UCS algorithm can
indeed find all $\ell$ indices. We now estimate an eigenvalue lower
bound on $A_{\ell}$. Let $\lambda^{(t)}$, $\widehat{\lambda}^{(t)}$
and $\lambda_j^{(t)}$ represent the values of $\lambda$,
$\widehat{\lambda}$ and $\lambda_j$, respectively, determined in
iteration $t$.  Then note that by the definitions of $\lambda$ and
$\widehat{\lambda}$ we have
\begin{eqnarray}
\lambda^{(0)}<\widehat{\lambda}^{(0)}\le\lambda^{(1)}<\widehat{\lambda}^{(1)}\le\cdots\le\lambda^{(\ell - 1)}<\widehat{\lambda}^{(\ell - 1)}.\nonumber
\end{eqnarray}
Define the following quantity and functions:
\begin{eqnarray}
\widehat{T} &\stackrel{def}=& T \left( 1 - \widehat{\lambda}^{(\ell-1)} \right) \label{def:T}, \\
g(t) \stackrel{def}= \frac{\ell\left( 1 - \frac{n}{\widehat{T}} \right)}{m-t+\widehat{T}-n}, \hspace{0.4cm}  &\text{and}& \hspace{0.6cm} F(\widehat{T}) \stackrel{def}= \frac{\ell \left(1-\frac{n}{\widehat{T}}\right)}{m-\frac{\ell -
1}{2}+\widehat{T}-n}-\frac{n}{\widehat{T}}. \nonumber
\end{eqnarray}
To bound $\lambda_{\text{min}}\left(A_{\ell}\right)$, we first establish a recurrence relation on $\widehat{\lambda}^{(\ell-1)}$. 
\begin{lemma} \label{lem:tech}
After the last iteration of the UCS algorithm, we have
\[ \widehat{\lambda}^{(\ell-1)} \geq \left( 1 - \widehat{\lambda}^{(\ell-1)} \right) \left[ \frac{1}{\ell} \sum_{t=0}^{\ell-1} g\left( t \right) - \frac{n}{\widehat{T}} \right].\]
\end{lemma}
\proof
Remember that ${\displaystyle T = \text{\bf tr} \left( A- \lambda^{(t)}I \right)^{-1} =  \sum_{j=1}^n\frac{1}{\lambda^{(t)}_j-\lambda^{(t)}}}$, and note that 
\begin{eqnarray}
{\displaystyle \frac{1-\lambda_j^{(t)}}{\lambda_j^{(t)}-\lambda^{(t)}} = \frac{1-\lambda^{(t)}}{\lambda_j^{(t)}-\lambda^{(t)}} + \frac{\lambda^{(t)}-\lambda_j^{(t)}}{\lambda_j^{(t)}-\lambda^{(t)}} = \frac{1-\lambda^{(t)}}{\lambda_j^{(t)}-\lambda^{(t)}} - 1. \label{eqn:part}}
\end{eqnarray}
The equation $f\left( \widehat{\lambda}^{(t)} \right) = 0$ gives
\begin{eqnarray}
{\displaystyle \left(\widehat{\lambda}^{(t)}-\lambda^{(t)}\right)\left(m-t+\sum_{j=1}^n\frac{1-\lambda_j^{(t)}}{\lambda_j^{(t)}-\lambda^{(t)}}\right)=\frac{\sum_{j=1}^n\frac{1-\lambda_j^{(t)}}{\left(\lambda_j^{(t)}-\lambda^{(t)}\right)\left(\lambda_j^{(t)}-\widehat{\lambda}^{(t)}\right)}}{\sum_{j=1}^n\frac{1}{\left(\lambda_j^{(t)}-\lambda^{(t)}\right)\left(\lambda^{(t)}_j-\widehat{\lambda}^{(t)}\right)}}}.\nonumber
\end{eqnarray}
Applying equation (\ref{eqn:part}) to both sides:
\begin{eqnarray*}
&&{\displaystyle \left(\widehat{\lambda}^{(t)}-\lambda^{(t)}\right)\left(m-t+\left(1-\lambda^{(t)}\right)T-n\right)}\\
&& {\displaystyle\hspace{1cm}=1-\lambda^{(t)}-\frac{\sum_{j=1}^n\frac{1}{\lambda_j^{(t)}-\widehat{\lambda}^{(t)}}}{\sum_{j=1}^n\frac{1}{\left(\lambda_j^{(t)}-\lambda^{(t)}\right)\left(\lambda^{(t)}_j-\widehat{\lambda}^{(t)}\right)}} }\\
&& {\displaystyle \hspace{1cm}\ge1-\lambda^{(t)}-\frac{n \left(\max_{j^*} \frac{1}{\lambda^{(t)}_{j^*}-\lambda^{(t)}} \right)}{ \left(\max_{j^*} \frac{1}{\lambda^{(t)}_{j^*}-\lambda^{(t)}} \right) \left(\sum_{j=1}^n\frac{1}{\lambda^{(t)}_j-\lambda^{(t)}}\right)} } \\
&& {\displaystyle \hspace{1cm}=1-\lambda^{(t)}-\frac{n}{T}.}
\end{eqnarray*}
Since 
\[{\displaystyle \left( \widehat{\lambda}^{(t-1)} - \lambda^{(t)} \right) \leq 0, \quad \mbox{and} \quad \left(\widehat{\lambda}^{(t)}-\lambda^{(t)} \right) \geq \frac{1-\lambda^{(t)}-\frac{n}{T}}{m-t+\left(1-\lambda^{(t)}\right)T-n},}\]
we have
\begin{eqnarray}
{\displaystyle \widehat{\lambda}^{(\ell - 1)}}&\ge& {\displaystyle \widehat{\lambda}^{(\ell - 1)} + \sum_{t=1}^{\ell - 1} \overbrace{\left( \widehat{\lambda}^{(t-1)} - \lambda^{(t)} \right)}^{\leq 0} -\lambda^{(0)} + \lambda^{(0)} }\nonumber  \\
&=& {\displaystyle \sum_{t=0}^{\ell - 1}\left(\widehat{\lambda}^{(t)}-\lambda^{(t)}\right)+\lambda^{(0)}}\nonumber\\
&\ge&{\displaystyle \sum_{t=0}^{\ell - 1}\frac{1-\lambda^{(t)}-\frac{n}{T}}{m-t+\left(1-\lambda^{(t)}\right)T-n}-\frac{n}{T}} \nonumber \\
&\ge&{\displaystyle\sum_{t=0}^{\ell - 1}\frac{1-\widehat{\lambda}^{(\ell - 1)}-\frac{n}{T}}{m-t+\left(1-\widehat{\lambda}^{(\ell - 1)}\right)T-n}-\frac{n}{T}}. \label{eqn:dec}
\end{eqnarray}
Inequality (\ref{eqn:dec}) follows by noting that the
terms in the sum are decreasing in $\lambda^{(t)}$.  The final
substitution is necessary because solving the preceding recurrence
relation is impractical.  To further simplify calculations, we define
\begin{eqnarray}
\widehat{T}:=T\left(1-\widehat{\lambda}^{(\ell - 1)}\right).\nonumber
\end{eqnarray}
Therefore,
\begin{eqnarray}
\widehat{\lambda}^{(\ell - 1)}&\ge&\left(1-\widehat{\lambda}^{(\ell - 1)}\right)\left[\sum_{t=0}^{\ell - 1}\frac{1-\frac{n}{\widehat{T}}}{m-t+\widehat{T}-n}-\frac{n}{\widehat{T}}\right]\nonumber\\
&=&\left(1-\widehat{\lambda}^{(\ell - 1)}\right)\left[ \frac{1}{\ell} \sum_{t=0}^{\ell - 1} g(t)-\frac{n}{\widehat{T}}\right].\hspace{2 em}\endproof\nonumber
\end{eqnarray}

Next, to demonstrate the effectiveness of the algorithm, we derive a lower bound for $\lambda_n$ after $\ell$ iterations.  This analysis will
involve selecting an appropriate $T$ to maximize the lower bound.
\begin{lemma} \label{lem:lowbound}
If $\widehat{T} > n$, then 
\[ {\displaystyle \lambda_{\min} \left( A_{\ell} \right) \geq \frac{F \left( \widehat{T} \right)}{1+F \left( \widehat{T} \right)} }.\]
\end{lemma}
\proof A key observation is that $g(t)$ is \emph{strictly} convex in $t$,
which is easily verified by showing that the second derivative
$\frac{d^2g}{dt^2} (t)$ is positive by our assumptions that
$\widehat{T} > n$ and $m \geq \ell > t$. Next, we apply Jensen's
Inequality for discrete sums~\cite{zorich} to the recurrence relation in Lemma \ref{lem:tech}:
\begin{eqnarray}
{\displaystyle \widehat{\lambda}^{(\ell - 1)}}&\ge&{\displaystyle   \left(1-\widehat{\lambda}^{(\ell - 1)}\right) \left[\frac{1}{\ell}\sum_{t=0}^{\ell - 1}\frac{\ell \left(1-\frac{n}{\widehat{T}}\right)}{m-t+\widehat{T}-n}-\frac{n}{\widehat{T}}\right]}\nonumber\\
&>& {\displaystyle \left(1-\widehat{\lambda}^{(\ell - 1)}\right) \left[\frac{\ell \left(1-\frac{n}{\widehat{T}}\right)}{\frac{1}{\ell}\sum_{t=0}^{\ell - 1} m-t+\widehat{T}-n}-\frac{n}{ \widehat{T}}\right] } \nonumber\\
&=& {\displaystyle \left(1-\widehat{\lambda}^{(\ell - 1)}\right) \left[\frac{\ell \left(1-\frac{n}{\widehat{T}}\right)}{m-\frac{\ell - 1}{2}+\widehat{T}-n}-\frac{n}{\widehat{T}}\right]}\nonumber\\
&=&{\displaystyle \left(1-\widehat{\lambda}^{(\ell - 1)}\right)F\left(\widehat{T}\right)}.\nonumber
\end{eqnarray}
Along with $\lambda_n > \widehat{\lambda}$ from Lemma \ref{lemma:1},  this finally leads to 
\begin{eqnarray}
{\displaystyle \lambda_{\min} = \lambda_n > \widehat{\lambda}^{(\ell - 1)} > \frac{F\left(\widehat{T}\right)}{1+F\left(\widehat{T}\right)}.}\label{eqn:lF}
\hspace{2 em}\endproof
\end{eqnarray} 
The expression on the right-hand side of~(\ref{eqn:lF}) is monotonically increasing
in $F$. So, maximizing $F\left(\widehat{T}\right)$ will also
maximize the lower bound on $\widehat{\lambda}^{(\ell - 1)}$.
\begin{lemma} \label{lem:max}
The function $F(\widehat{T})$ is maximized at
\[ {\displaystyle \widehat{T}^* = \frac{n\left(m+\frac{\ell+1}{2}-n\right)+\sqrt{n \ell \left(m-\frac{\ell - 1}{2}\right)\left(m+\frac{\ell+1}{2}-n\right)}}{\ell - n} }.\]
\end{lemma}
\proof
Setting the derivative of $F\left(\widehat{T}\right)$ to zero:
$${\displaystyle \frac{dF}{d\widehat{T}}=\frac{\left(n-\ell \right)T^2 + 2n\left(m + \frac{\ell+1}{2}-n \right) T + n\left( m+\frac{\ell+1}{2}-n \right) \left( m-\frac{\ell - 1}{2}-n \right) }{\widehat{T}^2\left( m-\frac{\ell - 1}{2}-n+\widehat{T}\right)^2} = 0.}$$
Solving for the desired root:
$$\widehat{T}^*=\frac{n\left(m+\frac{\ell+1}{2}-n\right)+\sqrt{n\ell\left(m-\frac{\ell - 1}{2}\right)\left(m+\frac{\ell+1}{2}-n\right)}}{\ell - n}.$$
We see that $\widehat{T}^*$ is the global maximum on the region $\widehat{T} \in \left(n, \infty \right)$ via the first derivative test since $\frac{dF}{d\widehat{T}} > 0$ for $n < \widehat{T} < \widehat{T}^*$ and $\frac{dF}{d\widehat{T}} < 0$ for $\widehat{T}^* < \widehat{T}$.
\hspace{2 em}\endproof

We remark that combining ($\ref{def:T}$) and ($\ref{eqn:lF}$) implies that the UCS algorithm should choose 
$T = \widehat{T}^*\left(1+F\left(\widehat{T}^*\right)\right)$ for effective column selection. We are now ready to estimate $\lambda_{\min} \left( A_{\ell}\right)$.
\begin{theorem} \label{thm:lower}
If $T$ is chosen according to Lemma \ref{lem:max} in the UCS algorithm, then
\[{\displaystyle \lambda_{\min} \left( A_{\ell} \right) > \frac{1}{\kappa},} \]
where $\kappa$ is defined in~(\ref{Eqn:alpha}).
\end{theorem} 
\proof We wish to apply our choice of $\widehat{T}$ to Lemma \ref{lem:lowbound}. We satisfy the assumption
\[{\displaystyle  \widehat{T}^*=\frac{n\left(m+\frac{\ell+1}{2}-n\right)+\sqrt{n\ell\left(m-\frac{\ell - 1}{2}\right)\left(m+\frac{\ell+1}{2}-n\right)}}{\ell - n} \geq \frac{n\left(m-n\right)}{\ell - n} \geq n .}\]
Therefore, plugging $\widehat{T}^*$ into (\ref{eqn:lF}) of Lemma \ref{lem:lowbound}: 
\begin{eqnarray}
\lambda_{\min} \left( A_{\ell} \right)&>&\frac{F\left(\widehat{T}\right)}{1+F\left(\widehat{T}\right)}\nonumber\\
&=&\frac{(\ell - n)\widehat{T}-n\left(m+\frac{\ell+1}{2}-n\right)}{\widehat{T}\left(m-\frac{\ell - 1}{2}-n+\widehat{T}\right)+(\ell - n)\widehat{T}-n\left(m+\frac{\ell+1}{2}-n\right)}\nonumber\\
&=&\frac{(\ell - n)^2}{\left(\sqrt{n\left(m+\frac{\ell+1}{2}-n\right)}+\sqrt{\ell\left(m-\frac{\ell+1}{2}\right)}\right)^2+(\ell - n)^2}. \hspace{2 em}\endproof\nonumber
\end{eqnarray}

\subsection{Correctness of the Unweighted Column Selection Algorithm}\label{sec:33}

We are now in a position to prove Theorem \ref{thm:unweighted}. Our
arguments are similar to those of the weighted sparsifier algorithm in~\cite{ramanujansparse}.

{\em Proof of Theorem \ref{thm:unweighted}.} By
Proposition~\ref{prop:upp}, we only need to show $\frac{1}{\kappa} L_G
\preceq L_H$. Consider the SVD of $W_G^{1/2} B_G$ in equation~(\ref{PolarMat}), and let ${x}$ be any vector such that $y = \Sigma_G V_G x \not=0$. Then 

\begin{eqnarray*}
{\displaystyle  L_G} &=& {\displaystyle B_G^T W_G B_G = V_G^T \Sigma_G^2V_G} ,  \\
{\displaystyle  L_H} &=& {\displaystyle  B_G^T W_H B_G  =  B_G^T W_G^{1/2} \Pi_{\ell}^T\Pi_{\ell} W_G^{1/2} B_G} \\
                     &=& {\displaystyle V_G^T \Sigma_G \left( U_G \Pi_{\ell}^T\Pi_{\ell} U_G^T\right)\Sigma_G V_G} . \end{eqnarray*}
 
It follows that 
\begin{eqnarray}
{\displaystyle \frac{{x}^TL_H{x}}{{x}^TL_G{x}}} & = & 
\frac{x^T\left(V_G^T \Sigma_G \left( U_G \Pi_{\ell}^T\Pi_{\ell} U_G^T\right)\Sigma_G V_G \right)x}{x^T\left(V_G^T \Sigma_G^2V_G\right)x }\nonumber\\
&=&{\displaystyle \frac{{y}^TU_G\Pi_{\ell}^T\Pi_{\ell}U_G^T{y}}{{y}^T{y}}.} \label{Eqn:y}
\end{eqnarray} 
On the other hand, by construction we have 
\[{\displaystyle A_{\ell}  = \sum_{j\in\Pi_{\ell}} {u}_j{u}_j^T= U_G \Pi_{\ell}^T\Pi_{\ell} U_G^T.}\]
With equation~(\ref{Eqn:y}), the Courant-Fisher min-max property gives
\[ {\displaystyle \frac{{x}^TL_H{x}}{{x}^TL_G{x}} =
\frac{{y}^TU_G\Pi_{\ell}^T\Pi_{\ell}U_G^T{y}}{{y}^T{y}} \ge  \lambda_{\min} \left( A_{\ell} \right) > \frac{1}{\kappa}, } \]
where the last line is due to Theorem \ref{thm:lower}.
\hspace{2 em}\endproof

\section{Performance Comparison of UCS and Other Algorithms}
This section compares the bound (\ref{Eqn:alpha}) to bounds of other current methods.  

\subsection{Comparison with Twice-Ramanujan Sparsifiers}
Given a weighted graph $G = \left(V,E, w \right)$, the algorithm of \cite{ramanujansparse} produces a sparsified graph $H = \left(V,F, \widehat{w} \right)$, where $F$ is a subset of $E$ and $\widehat{w}$ contains new edge weights, such that 
\begin{equation}\label{Eqn:Hw}
{\displaystyle L_G \preceq L_{H} \preceq \left(\frac{\sqrt{d}+1}{\sqrt{d}-1}\right)^2 L_G ,}
\end{equation}
where the parameter $d$ is defined via the equation $\ell = \left\lceil
d\left(n-1\right)\right\rceil$. 

By choosing $d$ to be a moderate and dimension-independent constant,
equation (\ref{Eqn:Hw}) asserts that every graph $G = \left(V,E, w
\right)$ has a \emph{weighted} spectral sparsifier with a number of
edges linear in $|V|$. This strong result, nevertheless, is obtained by
allowing unrestricted changes in the graph weights.  Such changes may be undesirable, especially if $G$ is unweighted, and the UCS algorithm may be preferred.

To compare the effectiveness of these two types of sparsifiers, we 
simplify equation~(\ref{Eqn:alpha}):
\[{\displaystyle \frac{1}{\kappa} \approx
\frac{\left(\sqrt{d}-1\right)^2}{m/n+d/2+\left(\sqrt{d}-1\right)^2}. } 
\]
It follows that for ${\kappa} = \Theta(1)$, a
dimension-independent constant, we must choose $d = \Theta(m/n)$. This
is the price one must pay to retain the original weights. For $d \ll
m/n$, the UCS algorithm computes a sparsified graph with
a $\kappa$ that grows at most linearly with $m/n$.  The algorithm of \cite{ramanujansparse} runs in time $O\left(dn^3m\right)$, which is equivalent to UCS.

\subsection{Further Comparisons of Column Selection Algorithms}
The algorithm of \cite{ramanujansparse} has been generalized in \cite{nearoptcol} to a column selection algorithm for computing CX decompositions.  In this work, Boutsidis, Drineas, and Magdon-Ismail prove that, given row-orthonormal matrices $V_1^T=\begin{pmatrix}\vec{v}_1^1&\vec{v}_2^1&\cdots&\vec{v}_m^1\end{pmatrix} \in \mathbb{R}^{n \times m}$ and $V_2^T=\begin{pmatrix}\vec{v}_1^2&\vec{v}_2^2&\cdots&\vec{v}_m^2\end{pmatrix}\in \mathbb{R}^{(m-n) \times m}$ then for a given $n<\ell\le m$ there exist weights $s_i \ge 0$ with at most $\ell$ of them nonzero such that
\begin{eqnarray}
 \lambda_{n}\left(\sum_{i=1}^ms_i\vec{v}_i^1\left(\vec{v}_i^1\right)^T\right)\ \ \ge\ \ \left(1-\sqrt{\frac{n}{\ell}}\right)^2\label{weighted0}
 \end{eqnarray}
 and
 \begin{eqnarray}
 \lambda_1\left(\sum_{i=1}^ms_i\vec{v}_i^2\left(\vec{v}_i^2\right)^T\right)\ \ \le\ \ \left(1+\sqrt{\frac{m-n}{\ell}}\right)^2.\label{weighted1}
\end{eqnarray}
In the context of CX decompositions, $\left[\begin{matrix}V_1^T\\ V_2^T\end{matrix}\right]=V^T\in\mathbb{R}^{m\times m}$ is understood to be the loadings matrix of a data matrix $A$, i.e. $A=U\Sigma V^T$ is the SVD of $A$ (although the algorithm could be applied to other matrices for other applications).  Their work includes an algorithm for finding the weights, Deterministic Dual Set Spectral Sparsification (DDSSS).

\begin{theorem}\label{bnd:old}
Let $\Pi_{\ell}^{DDSSS}\in\mathbb{R}^{m\times\ell}$ denote a matrix that chooses the $\ell$ columns selected by the DDSSS algorithm.  The inequalities (\ref{weighted0}) and (\ref{weighted1}) imply
\begin{eqnarray}
\sigma^2_{\min}\left(V_1^T \Pi_{\ell}^{DDSSS}\right)&\ge&\frac{\left(\sqrt{\ell}-\sqrt{n}\right)^2}{\left(\sqrt{\ell}+\sqrt{m-n}\right)^2+\left(\sqrt{\ell}-\sqrt{n}\right)^2}\nonumber\\
&\stackrel{def}=&\frac{1}{\kappa_{\text{DDSSS}}}.\nonumber
\end{eqnarray}
\end{theorem}
\proof
We interpret these inequalities as a bound on $\lambda_n$ by first partitioning
$$V^T\Pi=\left(\begin{array}{cc} V_1 & V_1^\prime \\ V_2 & V_2^\prime\end{array}\right),$$
where $\Pi$ is a permutation matrix that orders the selected columns first. Then, using a CS decomposition \cite{loan}, we can write
\begin{eqnarray}
\left(\begin{array}{c} V_1 \\ \\ V_2\end{array}\right)&=&\left(\begin{array}{c}P_1\left(\begin{array}{cc} C & 0\end{array}\right)Q_1^T\\ \\ P_2\left(\begin{array}{cc} -S & 0\\ 0 & I \\ 0 & 0\end{array}\right)Q_2^T\end{array}\right),\nonumber
\end{eqnarray}
where $C$ and $S$ are diagonal matrices with non-negative entries such that $C^2+S^2=I$.  Furthermore, because $P_1$ and $Q$ are orthogonal, by inspection $C$ contains the singular values of $V_1$.  Hence
\begin{eqnarray}
\lambda_n&\ge&\sigma_{\min}^2\left(V_1\right)=\sigma_{\min}^2(C).\nonumber
\end{eqnarray}
Now let $W$ be a weight matrix, whose diagonal entries are $\sqrt{s_i}$, the weights from above.  Define
\begin{eqnarray}
\widehat{Q}&\stackrel{def}=&Q_1^T\left(WW^T\right)Q_1\nonumber\\
&\stackrel{def}=&\left(\begin{array}{cc}\widehat{Q}_{11} & \widehat{Q}_{12} \\ \widehat{Q}_{21} & \widehat{Q}_{22}\end{array}\right).\nonumber
\end{eqnarray}
Then
\footnotesize
\begin{eqnarray}
&&\left(V_2W\right)\left(V_1W\right)^\dagger\nonumber\\
&&\hspace{1cm}=V_2W\left(V_1W\right)^T\left(V_1WW^TV_1^T\right)^{-1}\nonumber\\
&&\hspace{1cm}=V_2\left(WW^T\right)V_1^T\left(V_1\left(WW^T\right)V_1^T\right)^{-1}\nonumber\\
&&\hspace{1cm}=V_2\left(WW^T\right)\left(P_1\left(\begin{array}{cc} C&0\end{array}\right)Q_1^T\right)^T\left(P_1\left(\begin{array}{cc} C&0\end{array}\right)Q_1^T\left(WW^T\right)Q_1\left(\begin{array}{c} C\\ 0\end{array}\right)P_1^T\right)^{-1}\nonumber\\
&&\hspace{1cm}=P_2\left(\begin{array}{cc} -S & 0 \\ 0 & I \\ 0 & 0\end{array}\right)\widehat{Q}\left(\begin{array}{c} C \\ 0\end{array}\right)P_1^TP_1\left(\left(\begin{array}{cc} C & 0\end{array}\right)\widehat{Q}\left(\begin{array}{c} C \\ 0\end{array}\right)\right)^{-1}P_1^T\nonumber\\
&&\hspace{1cm}=P_2\left(\begin{array}{cc} -S & 0 \\ 0 & I \\ 0 & 0\end{array}\right)\left(\begin{array}{cc}\widehat{Q}_{11} & \widehat{Q}_{12} \\ \widehat{Q}_{21} & \widehat{Q}_{22}\end{array}\right)\left(\begin{array}{c} C \\ 0 \end{array}\right)\left(C\widehat{Q}_{11}C\right)^{-1}P_1^T\nonumber\\
&&\hspace{1cm}=P_2\left(\begin{array}{cc} -S & 0 \\ 0 & I \\ 0 & 0\end{array}\right)\left(\begin{array}{c}\widehat{Q}_{11}C \\ \widehat{Q}_{21}C\end{array}\right)C^{-1}\widehat{Q}_{11}^{-1}C^{-1}P_1^T\nonumber\\
&&\hspace{1cm}=P_2\left(\begin{array}{c} -SC^{-1} \\ \widehat{Q}_{21}\widehat{Q}_{11}^{-1}C^{-1} \\ 0\end{array}\right)P_1^T.\nonumber
\end{eqnarray}
\normalsize
Therefore
\begin{eqnarray}
\sqrt{\frac{1-\sigma_{\min}^2(C)}{\sigma_{\min}^2(C)}}&=&\|SC^{-1}\|_2\nonumber\\
&\le&\|\left(V_2W\right)\left(V_1W\right)^\dagger\|_2\nonumber\\
&\le&\left(1+\sqrt{\frac{m-n}{\ell}}\right)\left(1-\sqrt{\frac{n}{\ell}}\right)^{-1}.\nonumber
\end{eqnarray}

Rearranging
\begin{eqnarray}
\sigma_{\min}^2(C)&\ge&\frac{\left(1-\sqrt{\frac{n}{\ell}}\right)^2}{\left(1+\sqrt{\frac{m-n}{\ell}}\right)^2+\left(1-\sqrt{\frac{n}{\ell}}\right)^2}\nonumber\\
&=&\frac{\left(\sqrt{\ell}-\sqrt{n}\right)^2}{\left(\sqrt{\ell}+\sqrt{m-n}\right)^2+\left(\sqrt{\ell}-\sqrt{n}\right)^2}.\hspace{2 em}\endproof\nonumber
\end{eqnarray}

\begin{corollary}
Let $\kappa_{\text{UCS}}$ be as defined in equation (\ref{Eqn:alpha}).  Then
$$\frac{1}{\kappa_{\text{UCS}}}>\frac{1}{\kappa_{\text{DDSSS}}}.$$
\end{corollary}
\proof
\footnotesize
\begin{eqnarray}
&&\frac{(\ell-n)^2}{\left(\sqrt{n\left(m+\frac{\ell+1}{2}-n\right)}+\sqrt{\ell\left(m-\frac{\ell+1}{2}\right)}\right)^2+(\ell-n)^2}\nonumber\\
&&\hspace{1cm}=\frac{(\sqrt{\ell}-\sqrt{n})^2}{\left(\frac{\sqrt{n\left(m+\frac{\ell+1}{2}-n\right)}+\sqrt{\ell\left(m-\frac{\ell+1}{2}\right)}}{\sqrt{\ell}+\sqrt{n}}\right)^2+(\sqrt{\ell}-\sqrt{n})^2}\nonumber\\
&&\hspace{1cm}\ge\frac{(\sqrt{\ell}-\sqrt{n})^2}{\left(\frac{\sqrt{n\left(m+\frac{\ell+1}{2}-n\right)}+\sqrt{\ell m}}{\sqrt{\ell}+\sqrt{n}}\right)^2+(\sqrt{\ell}-\sqrt{n})^2}\nonumber\\
&&\hspace{1cm}\ge\frac{(\sqrt{\ell}-\sqrt{n})^2}{\left(\frac{\sqrt{n\left(m+\ell-n\right)}+\sqrt{\ell m}}{\sqrt{\ell}+\sqrt{n}}\right)^2+(\sqrt{\ell}-\sqrt{n})^2}\nonumber\\
&&\hspace{1cm}\ge\frac{(\sqrt{\ell}-\sqrt{n})^2}{\left(\frac{\sqrt{n\left(m+\ell-n\right)}+\sqrt{\ell m-n\ell+\ell^2}}{\sqrt{\ell}+\sqrt{n}}\right)^2+(\sqrt{\ell}-\sqrt{n})^2}\nonumber\\
&&\hspace{1cm}\ge\frac{(\sqrt{\ell}-\sqrt{n})^2}{\left(\frac{\sqrt{nm-n^2}+\sqrt{n\ell}+\sqrt{\ell m-n\ell}+\sqrt{\ell^2}}{\sqrt{\ell}+\sqrt{n}}\right)^2+(\sqrt{\ell}-\sqrt{n})^2}\nonumber\\
&&\hspace{1cm}=\frac{(\sqrt{\ell}-\sqrt{n})^2}{\left(\frac{\left(\sqrt{m-n}+\sqrt{\ell}\right)\left(\sqrt{\ell}+\sqrt{n}\right)}{\sqrt{\ell}+\sqrt{n}}\right)^2+(\sqrt{\ell}-\sqrt{n})^2}\nonumber\\
&&\hspace{1cm}=\frac{(\sqrt{\ell}-\sqrt{n})^2}{\left(\sqrt{m-n}+\sqrt{\ell}\right)^2+(\sqrt{\ell}-\sqrt{n})^2}.\hspace{2 em}\endproof\nonumber
\end{eqnarray}
\normalsize
This suggests the UCS algorithm may find a better subset than the column selection algorithm in \cite{nearoptcol}. Observe that typically $m \gg \ell \geq n$. For the purpose of finding a well-conditioned subset of columns in $V_1^T\in\mathbb{R}^{n\times m}$, requiring the whole matrix $V^T\in\mathbb{R}^{m\times m}$ is computationally expensive. On the other hand, an even better subset can be obtained by applying the UCS algorithm directly to $V_1^T$, at considerable savings in computational time and memory usage.  This algorithm runs in time $O\left(\ell m\left(n^2+\left(m-\ell\right)^2\right)\right)\approx O\left(\ell m^3\right)$, far slower than UCS.

\section{A Numeric Example: Graph Visualization}
We test the UCS algorithm on the Autonomous systems \mbox{AS-733} dataset in \cite{snapnets}\footnote{File {\tt as19981229}}.  The data is undirected, unweighted, and contains 493 nodes and 1189 edges.  To visualize the data, nodes are plotted using coordinates determined by the force-directed Fruchterman-Reingold algorithm.  This algorithm treats the edges of a graph as forces (similar to springs), and perturbs node coordinates until the graph appears to be near an equilibrium state \cite{FruRei91}.

We apply the force-directed algorithm with two methodologies.  First, the force-directed algorithm is run on the whole graph to determine a fixed set of node coordinates.  Using these coordinates, the original graph is plotted with various sparsifiers in Figure \ref{fig:2}.  Second, we run the force-directed algorithm on each sparsifier to determine node coordinates for that sparsifier, and plot both the sparsifier and the original graph on these coordinates (Figure \ref{fig:3}).  While this requires rerunning the force-directed algorithm for each sparsifier, the algorithm converges faster because of the reduced number of edges.

\begin{figure}[t]
\centering
    \includegraphics[width=0.5\textwidth]{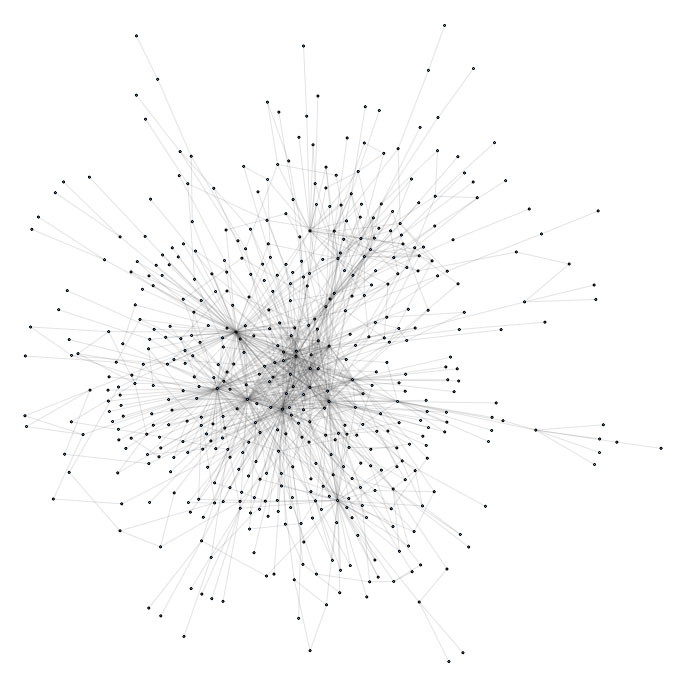}
\caption[]{Autonomous System Example: Original Graph}
\label{fig:1}
\vspace{-0.5cm}
\end{figure}

\afterpage{
\begin{figure}[t]
\centering
\subfigure[984 Edges]{
    \includegraphics[width=0.19\textwidth]{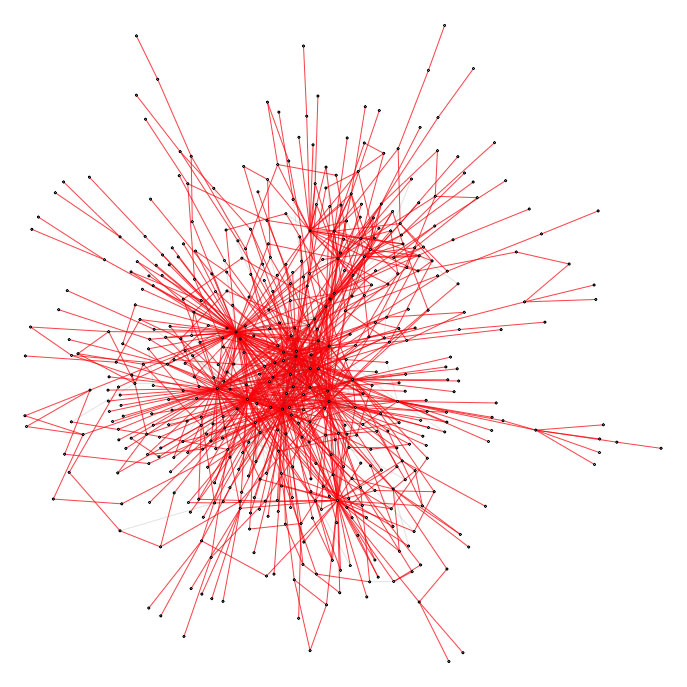}
}
\subfigure[738 Edges]{
    \includegraphics[width=0.19\textwidth]{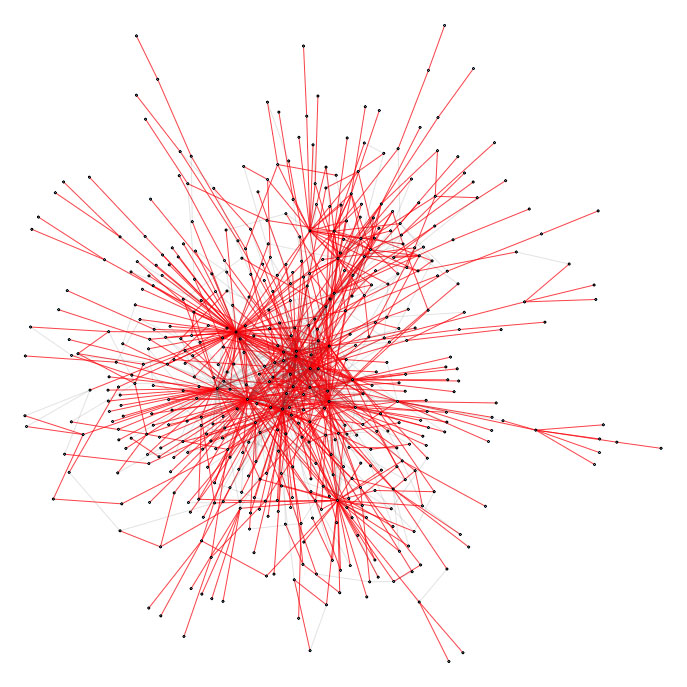}
}
\subfigure[615 Edges]{
    \includegraphics[width=0.19\textwidth]{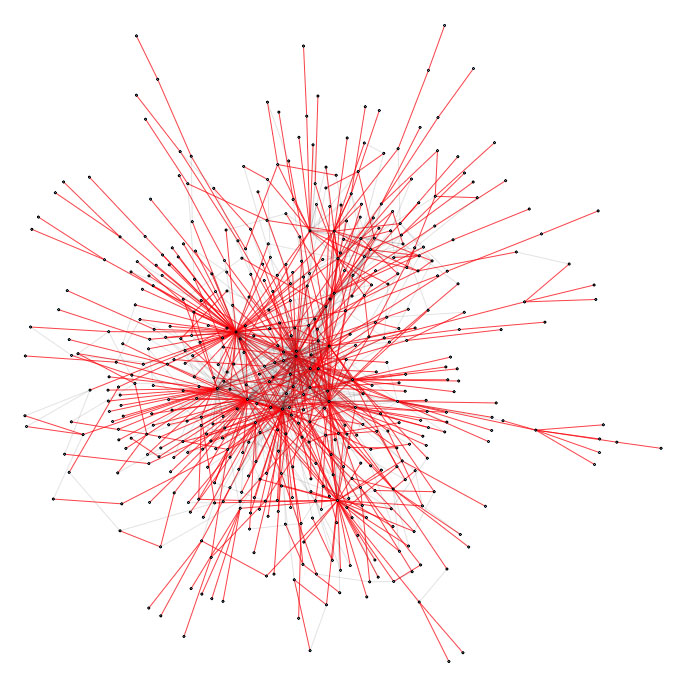}
}
\subfigure[Spanning Tree\protect\footnotemark]{
  \includegraphics[width=0.19\textwidth]{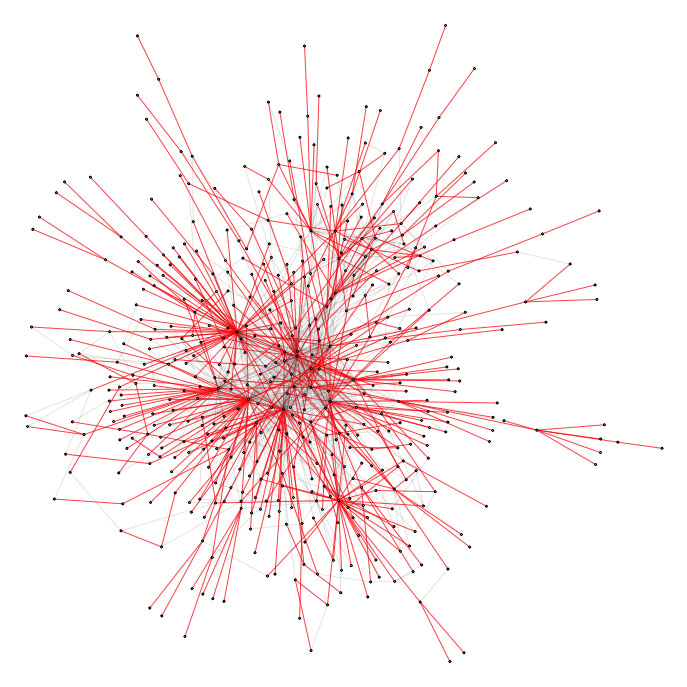}
}
\subfigure{
  \includegraphics[width=0.10\textwidth]{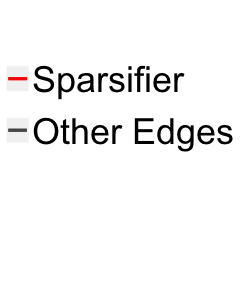}
}
\caption[]{Autonomous Systems Graph with Sparsifiers of Various Cardinalities (node coordinates calculated from whole graph)}
\label{fig:2}

\subfigure[984 Edges]{
    \includegraphics[width=0.19\textwidth]{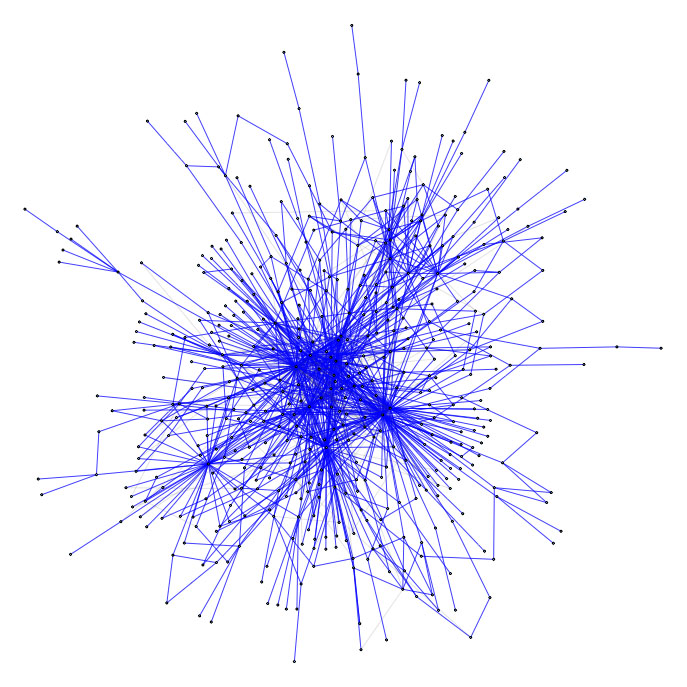}
}
\subfigure[738 Edges]{
    \includegraphics[width=0.19\textwidth]{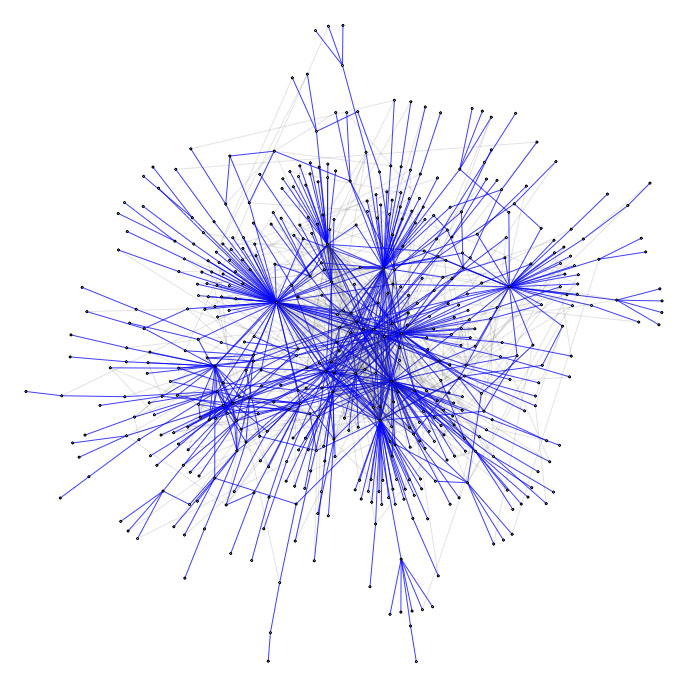}
}
\subfigure[615 Edges]{
    \includegraphics[width=0.19\textwidth]{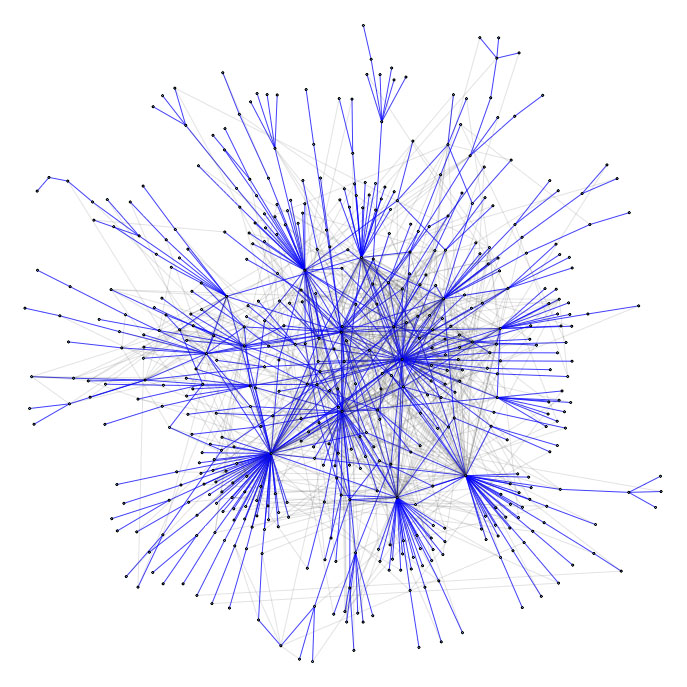}
}
\subfigure[{Spanning Tree\protect\footnotemark[3]}]{
  \includegraphics[width=0.19\textwidth]{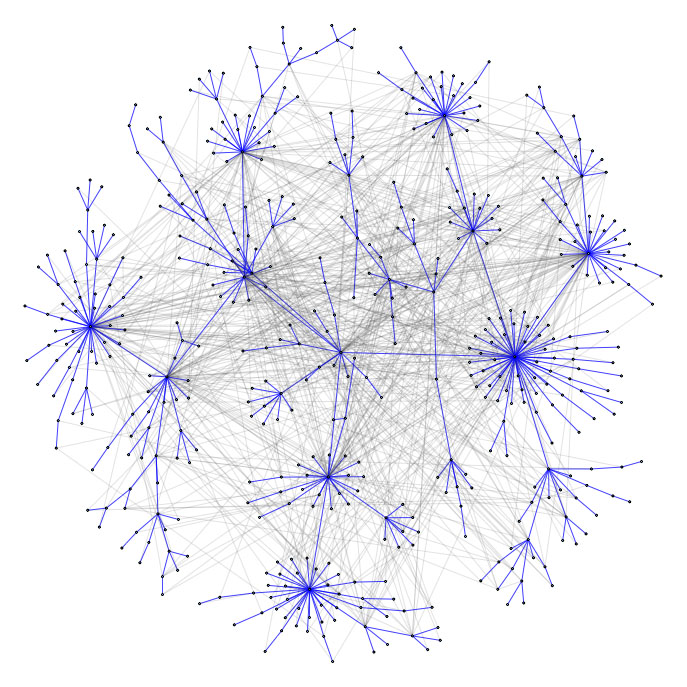}
}
\subfigure{
  \includegraphics[width=0.10\textwidth]{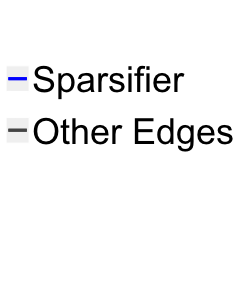}
}
\caption[]{Autonomous Systems Graph with Sparsifiers of Various Cardinalities (node coordinates recalculated for each sparsifier)}
\label{fig:3}

\subfigure[984 Edges]{
    \includegraphics[width=0.19\textwidth]{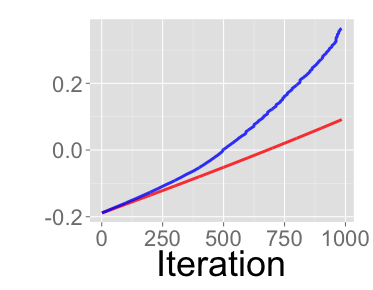}
}
\subfigure[738 Edges]{
    \includegraphics[width=0.19\textwidth]{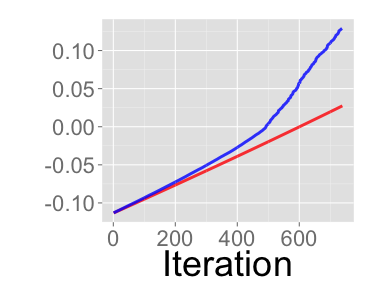}
}
\subfigure[615 Edges]{
    \includegraphics[width=0.19\textwidth]{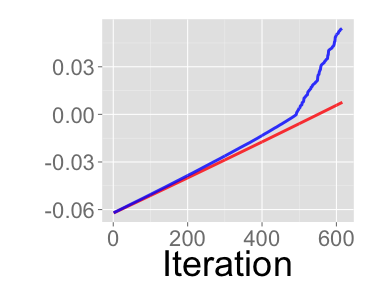}
}
\subfigure[493 Edges]{
  \includegraphics[width=0.19\textwidth]{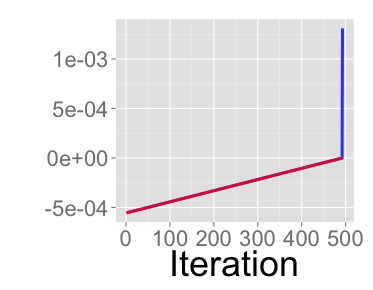}
}
\subfigure{
  \includegraphics[width=0.10\textwidth]{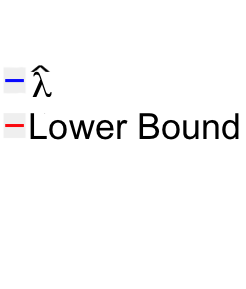}
}


\caption[]{Progress During Iteration and Theoretical Singular Value Lower Bound for Sparsifiers of Various Cardinalities}
\label{fig:4}

\vspace{-0.5cm}
\end{figure}
\footnotetext[3]{Calculated by running UCS algorithm with $\ell=493$ and omitting the final edge.  In general, a spanning tree for a connected graph can be found by selecting $\ell=n+1$ edges and removing an edge from a loop created by the UCS algorithm.}
}

Although the original graph can be considered sparse, visualization of the graph is difficult.  In Figure \ref{fig:1}, a few nodes are seen to have high degree, but little information is readily available about important edges in the graph or about how important nodes are related.  Figure \ref{fig:2} shows that plotting the sparsifier on the original graph provides incremental benefit.  The sparser graphs begin to highlight important nodes and important edges connecting them, but visualization remains difficult.  Rerunning the force-directed algorithm on the sparsifiers, nevertheless, evokes an easily interpretable structure, where important nodes, clusters, and important edges connecting clusters are readily visible (Figure \ref{fig:3}).  

\section{Relationship to the Kadison-Singer Problem} 
Let $p \geq 2$ be an integer, and let $ U = \left(u_1, \cdots,
u_m\right) \in \mathbb{R}^{n\times m}$ be a matrix that satisfies
\begin{equation}
\sum_{k=1}^n u_k u_k^T  =  I, \quad \mbox{and} \quad 
\|u_k\|_2  \leq  \delta, \quad \mbox{for}\quad k = 1, \cdots, m, \label{Eqn:delta}
\end{equation}
where $0 < \delta < 1$. Equation~(\ref{Eqn:delta}) implies that $U$ is
a row-orthonormal matrix and that each column of $U$ is uniformly
bounded away from $1$ in $2-$norm. Marcus {\em et al.} \cite{Marcus2014} show that there exists a partition 
\begin{equation}
{\displaystyle \mathcal{P} = \mathcal{P}_1 \cup \cdots \cup \mathcal{P}_p } \label{Eqn:Partition}
\end{equation}
of ${\displaystyle \{1, \cdots , n\}}$ such that 
\begin{equation}
{\displaystyle \left\|U\left(:,\mathcal{P}_k\right)\right\|_2 \leq \frac{1}{\sqrt{p}} + \delta, \quad \mbox{for}\quad k = 1, \cdots, p.}\nonumber
\end{equation}
When the graph $G$ is sufficiently dense, equation~(\ref{Eqn:Partition}) implies the existence of an unweighted graph sparsifier (see~Batson, {\em et al.}~\cite{ramanujansparse}) .

\section{Conclusion} We have presented an efficient algorithm for the construction of unweighted spectral sparsifiers for general weighted and unweighted graphs, addressing the open question of the existence of such graph sparsifiers for general graphs~\cite{ramanujansparse}.  Our algorithm is supported by strong theoretical spectral bounds.  Through numeric experiments, we have demonstrated that our sparsification algorithm can be an effective tool for graph visualization, and anticipate that it will prove useful for wide-ranging applications involving large graphs.  An important feature of our sparsification algorithm is the deterministic unweighted column selection algorithm on which it is based.  An open question is the existence of a larger lower spectral bound, either with the same $T$ or a new one.\\

\bibliographystyle{plain}
\bibliography{ucs_sparsifiers}

\begin{thebibliography}{10}

\bibitem{agm}
K.~J. Ahn, S.~Guha, and A.~McGregor.
\newblock Graph sketches: sparsification, spanners, and subgraphs.
\newblock In {\em PODS}, pages 5--14. ACM, 2012.

\bibitem{fasterSub}
H.~Avron and C.~Boutsidis.
\newblock Faster subset selection for matrices and applications.
\newblock {\em CoRR}, abs/1201.0127, 2012.

\bibitem{ramanujansparse}
J.~D. Batson, D.~A. Spielman, and N.~Srivastava.
\newblock Twice-ramanujan sparsifiers.
\newblock {\em SIAM J. Comput.}, 41(6):1704--1721, 2012.

\bibitem{journals/cacm/BatsonSST13}
J.~D. Batson, D.~A. Spielman, N.~Srivastava, and S.-H. Teng.
\newblock Spectral sparsification of graphs: theory and algorithms.
\newblock {\em Commun. ACM}, 56(8):87--94, 2013.

\bibitem{conf/stoc/BenczurK96}
A.~A. Bencz\'{u}r and D.~R. Karger.
\newblock Approximating s-t minimum cuts in ${O}(n^2)$ time.
\newblock In Gary~L. Miller, editor, {\em STOC}, pages 47--55. ACM, 1996.

\bibitem{nearoptcol}
C.~Boutsidis, P.~Drineas, and M.~Magdon-Ismail.
\newblock Near-optimal column-based matrix reconstruction.
\newblock {\em CoRR}, abs/1103.0995, 2011.

\bibitem{conf/soda/ChierichettiLP10}
F.~Chierichetti, S.~Lattanzi, and A.~Panconesi.
\newblock Rumour spreading and graph conductance.
\newblock In {\em SODA}, pages 1657--1663. SIAM, 2010.

\bibitem{DBLP:conf/stoc/ChristianoKMST11}
P.~Christiano, J.~A. Kelner, A.~Madry, D.~A. Spielman, and S.-H. Teng.
\newblock Electrical flows, laplacian systems, and faster approximation of
  maximum flow in undirected graphs.
\newblock In Lance Fortnow and Salil~P. Vadhan, editors, {\em STOC}, pages
  273--282. ACM, 2011.

\bibitem{FruRei91}
T.~Fruchterman and E.~Reingold.
\newblock Graph drawing by force-directed placement.
\newblock {\em Software--Practice {\&} Experience}, 21(11):1129--1164, 1991.

\bibitem{conf/innovations/KapralovP12}
M.~Kapralov and R.~Panigrahy.
\newblock Spectral sparsification via random spanners.
\newblock In Shafi Goldwasser, editor, {\em ITCS}, pages 393--398. ACM, 2012.

\bibitem{4495}
C.~Leiserson.
\newblock {Fat-trees: Universal Networks for Hardware-efficient
  Supercomputing}.
\newblock {\em IEEE Trans. Comput.}, 34(10):892--901, 1985.

\bibitem{snapnets}
J.~Leskovec and A.~Krevl.
\newblock {SNAP Datasets}: {Stanford} large network dataset collection.
\newblock \url{http://snap.stanford.edu/data}, October 2014.

\bibitem{loan}
C.~Van Loan.
\newblock Computing the cs and the generalized singular value decompositions.
\newblock {\em Numerische Mathematik}, 46, Issue 4:479--491, 1985.

\bibitem{Marcus2014}
A.~W. Marcus, D.~A. Spielman, and N.~Srivastava.
\newblock Interlacing families {II}: Mixed characteristic polynomials and the
  {Kadison-Singer} problem.
\newblock {\em CoRR}, abs/1306.3969, 2014.

\bibitem{conf/kdd/MathioudakisBCGU11}
M.~Mathioudakis, F.~Bonchi, C.~Castillo, A.~Gionis, and A.~Ukkonen.
\newblock Sparsification of influence networks.
\newblock In Chid Apt�, Joydeep Ghosh, and Padhraic Smyth, editors, {\em
  KDD}, pages 529--537. ACM, 2011.

\bibitem{ss1}
D.~A. Spielman and N.~Srivastava.
\newblock Graph sparsification by effective resistances.
\newblock {\em SIAM J. Comput.}, 40(6):1913--1926, 2011.

\bibitem{journals/corr/cs-DS-0310036}
D.~A. Spielman and S.-H. Teng.
\newblock Solving sparse, symmetric, diagonally-dominant linear systems in time
  ${O}(m^{1.31})$.
\newblock {\em CoRR}, cs.DS/0310036, 2003.

\bibitem{st1}
D.~A. Spielman and S.-H. Teng.
\newblock Nearly-linear time algorithms for graph partitioning, graph
  sparsiﬁcation, and solving linear systems.
\newblock In {\em {STOC'04}}, pages 81--90, 2004.

\bibitem{journals/corr/abs-cs-0607105}
D.~A. Spielman and S.-H. Teng.
\newblock Nearly-linear time algorithms for preconditioning and solving
  symmetric, diagonally dominant linear systems.
\newblock {\em CoRR}, abs/cs/0607105, 2006.

\bibitem{spectsim}
D.~A. Spielman and S.-H. Teng.
\newblock Spectral sparsification of graphs.
\newblock {\em CoRR}, abs/0808.4134, 2008.

\bibitem{zorich}
V.~A. Zorich.
\newblock {\em Mathematical Analysis I}.
\newblock Springer, Berlin, 2004.

\end{thebibliography}





\end{document}